\documentclass[11pt, journal, letterpaper, oneside, onecolumn]{IEEEtran}

\usepackage[cmex10]{amsmath}

\usepackage[all]{xy}

\usepackage{amsfonts}
\usepackage{amssymb}

\usepackage{mathrsfs}

\usepackage{graphicx}
\usepackage{enumerate}
\usepackage{url}
\usepackage{tikz}
\usetikzlibrary[topaths]
\usepackage[caption=false]{subfig}
\usepackage{comment}

\newcommand{\BB}{\mathbb{B}}

\newcommand{\JJ}{\mathcal{J}}
\newcommand{\II}{\mathcal{I}}
\newcommand{\KK}{\mathcal{K}}

\newcommand{\Fou}{\mathcal{F}}
\newcommand{\F}{\mathcal{F}}
\newcommand{\qed}{\hspace*{\fill}~{\rule{2mm}{2mm}}\par\endtrivlist\unskip}

\newcommand{\Z}{\mathbb{Z}}

\newcommand{\A}{\mathscr{A}}
\newcommand{\G}{\mathfrak{G}}

\newcommand{\J}{\mathcal{J}}

\newcommand{\D}{\mathcal{D}}

\newcommand{\PP}{\mathcal{P}}

\newcommand{\ZZ}{\mathcal{Z}}

\newcommand{\hZd}{\D(h)}

\newcommand{\om}{\omega}

\newcommand{\dl}{\delta}

\newcommand{\m}{\mathfrak{m}}
\newcommand{\n}{\mathfrak{n}}
\newcommand{\vm}{\bar{\mathfrak{m}}}
\newcommand{\MC}{\mathcal{L}}
\newcommand{\Prob}{\mathfrak{i}}

\newtheorem{lemma}{Lemma}
\newtheorem{theorem}{Theorem}
\newtheorem{thm8}{Theorem}
\newtheorem{corollary}{Corollary}
\newtheorem{proposition}{Proposition}

\newtheorem{conjecture}{Conjecture}

\title{Discrete Sampling: A graph theoretic approach to Orthogonal Interpolation}

\author{Aditya Siripuram, \quad William Wu \quad Brad Osgood}

\IEEEspecialpapernotice{Dedicated to the memory of Sivatheja Molakala}

\begin{document}

\maketitle
\begin{abstract}
We study the problem of finding unitary submatrices of the $N \times N$ discrete Fourier transform matrix, in the context of interpolating a discrete bandlimited signal using an orthogonal basis. This problem is  related to a diverse set of questions on idempotents on $\Z_N$ and tiling $\Z_N$. In this work, we establish a graph-theoretic approach and 
connections to the problem of finding maximum cliques. We identify the key properties of these graphs that make the interpolation problem tractable when $N$ is a prime power, and we identify the challenges in generalizing to arbitrary $N$. Finally, we investigate some connections between graph properties and the spectral-tile direction of the Fuglede conjecture.  \end{abstract}

\begin{IEEEkeywords} 
Discrete Fourier transforms, Interpolation, Perfect graphs, Circulant graphs
\end{IEEEkeywords}

\section{Introduction} \label{section:introduction}

\begin{figure}[htb]
\begin{center}
	
	\begin{tikzpicture}[transform shape,line width=0.01pt, scale = 0.7]
	
	\foreach \x in {0,...,15}{%
		\pgfmathparse{\x*22.5}
		\node[draw,circle,inner sep=0.05cm] (N-\x) at (\pgfmathresult:3.2cm) {\x};
	} 
	\foreach \x [count=\xi from 0] in {1,...,15}{%
		\foreach \y in {2,6,8,10,14}{%
			\pgfmathtruncatemacro\result{mod(\xi+\y, 16)}
			\path (N-\xi) edge[-] (N-\result);
		}
	\path (N-7) edge[thick, -] (N-15);
	\path (N-1) edge[thick, -] (N-9);
	\path (N-1) edge[thick, -] (N-7);
	\path (N-7) edge[thick, -] (N-9);
	\path (N-15) edge[thick, -] (N-9);
	\path (N-1) edge[thick, -] (N-15);
	}
	\end{tikzpicture}
	\label{fig:opening-example}
	\caption{For $\JJ = \{0,1,4,5\}$, we see that $\{1,7,9,15\}$ forms a clique, and hence an orthogonal sampling set}
	
\end{center}
\end{figure}
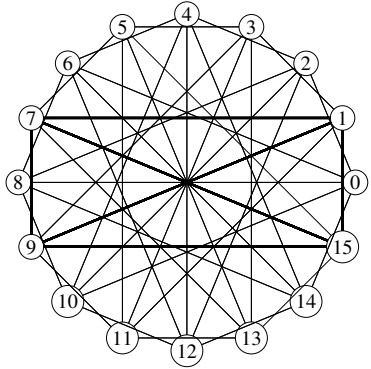

The  
simplest form of the question that we consider is this: 
\begin{itemize}
\item Which submatrices of the discrete Fourier transform are unitary up to scaling?
\end{itemize} 
For example, if from the $16 \times 16$ Fourier matrix we select rows $1,7,9,15$ and columns $0,1,4,5$ then the resulting $4 \times 4$ submatrix is unitary up to a factor of $4$. This is encoded in the graph of Fig \ref{fig:opening-example} and the corresponding clique shown in bold. 

Answers to the bulleted question involve the structure of convolution idempotents on $\mathbb{Z}_N$ (integers modulo $N$), tiling the integers \cite{farkas2006fuglede}, maximum cliques, and perfect graphs. As different as these areas might seem, what is most striking to us is that they intersect in the fundamental problem of sampling and interpolation for discrete signals.  This is the problem that first motivated us, and the  work here is a sequel to \cite{DS-1}. While we will refer to some of the results there we have tried to make the present paper self-contained.  

\subsection{Sampling, Interpolation, and Interpolating Bases} \label{section:sampling-and-interpolation}


 We need a few definitions to understand the connection to sampling and interpolation. Let $N$ be a positive integer and let $\omega_N=e^{2\pi i /N}$. For a discrete signal $f$ the Fourier transform and its inverse are
\[
\nonumber \Fou f(m) = \sum_{n=0}^{N-1} f(n) \omega_N^{-mn}, \quad
\nonumber \F^{-1} f(m) = \frac{1}{N}\sum_{n=0}^{N-1} f(n) \omega_N^{mn}.
\]
With this definition $\F\F^*=N I$ as matrices. Multiples of the identity also occur for the submatrices we look for  so, to keep from qualifying every such statement, from now on when we say ``unitary" we mean ``unitary up to scaling;" the multiple itself is not important.  Also, the particular way our problems arise makes it more natural to seek unitary submatrices of $\F^{-1}$ rather than of $\F$, but the principles are the same.  

We regard all discrete signals as mappings $f \colon \Z_N \longrightarrow \mathbb{C}$.  
 Likewise we always consider index sets to be subsets of $\Z_N$ so that algebraic operations on their elements are taken modulo $N$.  For $\JJ \subseteq \Z_N$ let 
\[
\BB^\JJ = \{f \colon \Fou f(n) =0, n \in \Z_N\setminus \JJ\}.
\]
In words, $\BB^\JJ$ is the $|\JJ|$-dimensional subspace of signals whose frequencies are zero off $\JJ$; there may be additional zeros but there are at least these. We do \emph{not} assume that $\J$ is a band of contiguous indices (mod $N$), but for short we still employ the term \textit{bandlimited} for a signal in $\BB^\JJ$. 

If $f\in \BB^\JJ$ for some $\JJ$ we ask if all values of $f$ can be interpolated from the sampled values $f(i)$ with $i$ drawn from an index set $\II \subseteq \Z_N$. If so, we say that $\II$ \emph{solves the sampling problem} for $\BB^\JJ$ and we call $\II$ a \emph{sampling set}. 
Associated with a sampling set is an \emph{interpolating basis} $\{u_i \colon i \in \II\}$ of $\BB^\JJ$ for which
\[
f = \sum_{i\in \II} f(i) u_i
\]
for any $f \in \BB^\JJ$.
The point of an interpolating basis, as opposed to any other basis of $\BB^\JJ$, is that the coefficients are the components of $f$ with respect to the natural basis of $\mathbb{C}^N$. A basis of $\BB^\JJ$ is an interpolating basis if and only if it satisfies
\begin{equation} \label{eq:ib-property}
u_i(j) = \dl_{ij}, \quad i,j \in \II.
\end{equation}
See  \cite{DS-1} for further general properties of interpolating bases. Every $\BB^\JJ$ has an interpolating basis but \emph{not} every $\BB^\JJ$ has an \emph{orthogonal} interpolating basis, and this is the starting point of our study.
 
It is easy to give a criterion for the sampling problem to have a solution, and this relates sampling to submatrices of the Fourier matrix. First, in general, for an index set $\KK
\subseteq \Z_N$ let $E_\KK$ be the $N \times |\KK|$ matrix whose columns are the natural basis vectors of $\mathbb{C}^N$ indexed, in order, by $\KK$. Then $E_\II^\textsf{T} f$ and $E_\JJ^\textsf{T}(\Fou f)$ are respectively the column vectors of the samples of $f$ on $\II$ and the samples of $\Fou f$ on $\JJ$. The matrix $E_\II^\textsf{T}\Fou^{-1} E_\JJ$ is the submatrix of $\Fou^{-1}$ with rows indexed by $\II$ and columns indexed by $\JJ$. Then as shown in 
\cite{DS-1}: 
\begin{theorem} \label{theorem:sampling-problem-solution} $\II$ solves the sampling problem for   $\BB^\JJ$ if and only if the submatrix $E_\II^\textsf{T}\Fou^{-1} E_\JJ$ of $\Fou^{-1}$ is invertible.
\end{theorem}
In particular $|\II|=|\JJ|$.  One has 
\begin{equation}   \label{eq:sampling-recovery}
E_\II^\textsf{T}f = (E_\II^\textsf{T}\Fou^{-1}E_\JJ)(E_\JJ^\textsf{T}\Fou f),
\end{equation}
and the interpolation formula  
\[
f = \Fou^{-1}E_\JJ(E_\II^\textsf{T}\Fou^{-1} E_\JJ)^{-1}E_\II^\textsf{T}f = U(E_\II^\textsf{T}f),
\]
expressing all the values of $f$ in terms of the sampled values $E_\II^\textsf{T}f$ on $\II$. The columns $\{u_{i_1},u_{i_2},\dots \}$  of $U$ are  an  {interpolating basis} of $\BB^\JJ$.

In \cite{DS-1} we were concerned primarily with \emph{universal sampling sets}. These are the index sets $\II$ for which $E_\II^\textsf{T} \Fou^{-1} E_\JJ$ is invertible {\emph{for every} index set $\JJ$  
of the same size as $\II$. So $\II$ is universal if having chosen rows in $\Fou^{-1}$ according to $\II$, \emph{any} choice of $|\II|$ columns results in an invertible submatrix. In terms of the sampling problem, $\II$ is a sampling set for {any} space $\BB^\JJ$, so for a universal sampling set the interpolating basis for $\BB^\JJ$ changes with $\JJ$, but {where} to sample does not. 

Theorem \ref{theorem:sampling-problem-solution} only goes so far. We need to be able to solve \eqref{eq:sampling-recovery} feasibly even in the presence of noise or numerical errors, and  universality does not guarantee such \emph{stable recovery}. For this we need the matrix $E_\II^\textsf{T}\F^{-1}E_\JJ$ to be well conditioned,  along the lines of stability conditions imposed by other models,  \cite{landau:density}, \cite{Bass04randomsampling}. For {stable} recovery the sampling set $\II$ needs to be such that the energy in the samples is a a non-negligible fraction of the energy in the entire signal,
 \[
   ||E_\II^{\textsf{T}}f|| \geq \alpha ||f||,\quad  f \in \BB^\JJ, \alpha > 0,
 \]
for a positive constant $\alpha$. This constraint is typically imposed for discrete (not necessarily timelimited) signals, \cite{Bass04randomsampling}. In the case we are investigating (timelimited discrete signals) we would typically want $\alpha$ to be as large as possible, uniformly over  $\BB^\JJ$. In other words, 
 \[
  \min_{ f \in \BB^\JJ} \frac{\|E_\II^\textsf{T}f\|}{\| f \|} \geq \alpha, \quad f \ne 0,
 \] 
 is equivalent to a condition on the norm of the Fourier submatrix,
\[
  \|E_\II^\textsf{T}\F^{-1}E_\JJ\| \geq \alpha,
\]
which requires that all the singular values of $E_\II^\textsf{T}\F^{-1}E_\JJ$ be at least $\alpha$.

From this point of view, the opening question of this paper thus identifies the extreme case when $E_\II \F^{-1} E_\JJ$ has the largest possible norm, and we want to know: 
\begin{itemize}
\item Given the frequency set $\JJ$, is it possible to find $\II$ such that  $E_\II^\textsf{T} \F^{-1} E_\JJ$ is unitary?   
\end{itemize}
If so we say that $(\II,\JJ)$ are a \emph{unitary pair}; rows of $\F^{-1}$ chosen according to $\II$ and columns according to $\JJ$, like the example given in the opening paragraph. Of course, since $(1/N)\F^{-\textsf{T}} = \F^{-1}$ we have that $E_\II^\textsf{T} \F^{-1} E_\JJ$ is unitary if and only if  $E_\JJ^\textsf{T} \F^{-1} E_\II$ is, so being a unitary pair is symmetric in $\II$ and $\JJ$. 

It is equivalent to the preceding question to ask:
\begin{itemize}
\item Does $\BB^\JJ$ have an orthogonal interpolating basis? 
\end{itemize}
If the answer is yes then the set $\II$ that indexes the orthogonal basis  (and so determines {where to sample}) will be called an \emph{orthogonal sampling set}. Our emphasis is on finding orthogonal sampling sets.  

The only $\BB^\JJ$ having an \emph{orthonormal} interpolating basis is all of $\mathbb{C}^N$,
while  proper subspaces $\BB^\JJ$ having an orthogonal interpolating basis cannot be too big:
\begin{proposition} \label{prop:N/2}
If $\BB^\JJ$ has an orthogonal interpolating basis then $|\JJ| \le N/2$.
\end{proposition}

We can express the symmetry of $\II$ and $\JJ$ as a unitary pair, or as orthogonal sampling sets, as:
\begin{proposition}
\label{prop:interchange}
 An index set $\II$ is an orthogonal sampling set for $\BB^\JJ$ if and only if $\JJ$ is an orthogonal sampling set for $\BB^\II$.	
\end{proposition}

See \cite{DS-1} for these results. The condition in Proposition \ref{prop:N/2} is not sufficient, but necessary and sufficient conditions for $\BB^\JJ$ to have an {orthogonal} interpolating basis \emph{in the case that $N$ is a prime power}, are known in the (mathematics) literature, though not in this context nor in this terminology. We will see one such result as Theorem \ref{thm:OIS} in Section \ref{section:idempotents}. 
The restriction that $N$ be a prime power also came up, in different ways, in \cite{DS-1}. Going beyond prime powers is an unmistakeable challenge. 


Our approach to the problem opens with an important connection between orthogonal sampling sets and the algebraic structure of the zeros of an idempotent $h=\F^{-1}1_\JJ$ on $\Z_N$ that comes from a given $\BB^\JJ$, where $1_\JJ$ is the indicator function of $\JJ$. We then introduce the \emph{difference graph} of an idempotent, recasting the problem of finding orthogonal sampling sets in graph theoretic terms as a search for maximum cliques, the key contribution of this paper. This is of more than theoretical interest because we show, in section \ref{section:perfect-graphs} that when $N$ is a prime power the difference graphs we consider are perfect graphs, and hence the problem of finding a maximum clique is tractable. We provide some insights into the challenges in generalizing such a result when $N$ is not a prime power, in section \ref{section:perfectness-abritrary-N}. We also investigate the clique, chromatic and Lov{\'a}sz numbers of these graphs in Section \ref{sec:other-graph-parameters} and their relationship to the Fuglede conjecture. We conclude with some interesting open problems.




We are happy to thank many colleagues for their interest, in particular Maria Chudnovsky, Sinan Gunturk, and Mark Tygert. We also thank Sivatheja Molakala, a close friend of the first author, and we dedicate this paper to his memory. 
\subsection{Related work on compressed sensing}
A somewhat related problem comes up in construction of Fourier sub-matrices with the restricted isometry property. The problem in this context is to find a set of rows $\II$ such that 
\[
\|(1-\delta)x\| \leq \| E_\II^\textsf{T}\F x\| \leq \|(1+\delta)x\|
\]
\emph{ for any } $N-$length vector $x$ that has at most $k<|\II|$ entries \cite{candes2006stable}. Thus, here the objective is to find a set of rows $\II$ such that \emph{regardless} of the choice of $k$ columns, the resulting matrix is \emph{approximately unitary}. Most of the initial constructions of such $\II$ were random \cite{candes2006near}, \cite{rudelson2008sparse}. Deterministic construction of such $\II$ has been harder, but significant developments were made in \cite{xia2005achieving}, \cite{haupt2010restricted} and \cite{xu2015compressed}, for example. In contrast, for the problem currently under investigation; we are given a set of columns $\JJ$, and we wish to find a set of rows $\II$ such that the resulting square submatrix is unitary.

\section{Idempotents on $\mathbb{Z}_N$ and their Zero-sets}  
\label{section:idempotents}

There is  a natural, one-to-one correspondence between index sets and idempotents for convolution, and properties of one can be used to study properties of the other.
 Given $\JJ\subseteq\Z_N$ let $1_\JJ\colon \Z_N \longrightarrow \{0,1\}$ be its indicator function.  Since $1_\JJ1_\JJ = 1_\JJ$ the function
\begin{equation} \label{eq;h}
 h_\JJ= \F^{-1} 1_\JJ
\end{equation} 
 is an idempotent, and  $h_\JJ\in \BB^\JJ$ by definition of $\BB^\JJ$. We write simply $h$ if the set $\JJ$ is clear from the context.  Conversely, if an idempotent $h$ is given then its Fourier transform, having values in $\{0,1\}$,  is the indicator function for an index set.

The following lemma opens the way to a very diverse set of phenomena.  A similar result, in a much different context, holds for discrete signals in $l^2(\mathbb{Z})$; see \cite{farkas2006fuglede}.

\begin{lemma} \label{lem:oib}
An index set $\II\subseteq \Z_N$ is an orthogonal sampling set for $\BB^\JJ$ if and only if $|\II| = |\JJ|$ and  $h(i_1-i_2)=0$ for $i_1,i_2 \in \II$, $i_1 \ne i_2$. 
\end{lemma}

\begin{IEEEproof}
Recall that a matrix is circulant if and only if it is diagonalized by the Fourier transform. Consider the matrix
\begin{align*}
H &= (E_\II^\textsf{T}\F^{-1} E_\JJ)(E_\II^\textsf{T}\F^{-1} E_\JJ)^{*} \\
&= E_\II^\textsf{T}\left(\F^{-1} E_\JJ E_\JJ^\textsf{T}(\F^{-1})^*\right)E_\II.
\end{align*}
Since  $E_\JJ E_\JJ^\textsf{T}$ is a diagonal matrix with the diagonal equal to $1_\JJ$, it follows that the matrix $ G = \F^{-1} E_\JJ E_\JJ^\textsf{T}(\F^{-1})^*$ is circulant, with first column $Nh$.
Hence the matrix $H$ is a submatrix of the circulant matrix $G$, with rows and columns both indexed by $\II$; in other words, the entires of the matrix $H$ are given by $N h(i - j)$ where  $i,j \in \II$. Now if  $E_\II^\textsf{T}\F E_\JJ$ is unitary, then $H$ is diagonal, which implies
\[
 h(i_1 - i_2) = 0 \text{ whenever } i_1 \neq i_2 \in \II.
\]
\end{IEEEproof}

In geometric terms, $h$ defines the orthogonal projection  $K\colon \mathbb{C}^N \longrightarrow  \BB^\JJ$ via
\[
Kv = h*v.
\]
The orthogonal complement to $\BB^\JJ$ is $\BB^{\JJ^c}$ where $\JJ^c=\Z_N\setminus \JJ$ is the complement of $\JJ$. 

The proof of Lemma \ref{lem:oib} is in keeping with the question we asked at the beginning of the paper, but it is worth pointing out an essentially equivalent approach. Let  $\tau:\Z_N\longrightarrow \Z_N$ be the shift $\tau(n) = n-1$, and write
\[
\tau^kh(n) = h(n-k).
\]
It is straightforward to check that the inner product of two shifted $h$'s is
\[
(\tau^ih,\tau^jh) = (h*h)(i-j) = h(i-j).
\]
Any shift of $h$ is also in $\BB^\JJ$ and thus an index set $\II\subseteq \Z_N$ having the property that $h(i_1-i_2)=0$ for $i_1,i_2 \in \II$, $i_1 \ne i_2$ determines a set of $|\II|$ orthogonal vectors in $\BB^\JJ$. When $\II$ is an orthogonal sampling set we normalize as in \eqref{eq:ib-property} to then obtain an orthogonal interpolating basis for $\BB^\JJ$ given by
\begin{equation} \label{eq:oib}
\{{h(0)}^{-1}\tau^ih\colon i \in \II\}.
\end{equation}
This is the recipe for turning an orthogonal sampling set into an orthogonal interpolating basis. All vectors in an orthogonal interpolating basis have length $(N/|\JJ|)^{1/2}$. 

In view of Lemma \ref{lem:oib} we introduce the \emph{zero-set} of the idempotent $h$ associated with a given $\JJ$,
\[
\ZZ(h) = \{n\in \Z_N \colon h(n)  = 0\}\,.
\]
Note that 
\[
h(0) = |\JJ|/N = \|h\|^2,
\]
so in particular  $0$ is never in $\ZZ(h)$. We also observe the symmetry relation 
\begin{equation} \label{eq:h-symmetry}
h(-m) = \overline{h(m)},
\end{equation}
implying  that both $m$ and $-m$ either are or are not in $\ZZ(h)$.

 It is  important for our work that the zero-set has a very particular algebraic structure. Let
\begin{equation} \label{eq:D_N}
\D_N = \{a: a\mid N \,\text{and} \, 1 \le a <N\}
\end{equation}
be the set of divisors of $N$ that are $<N$. Then:
\begin{lemma} \label{lem:h-null}
If $h$ is an idempotent then its zero-set $\ZZ(h)$ is the disjoint union 
\[
 \ZZ(h) = \bigcup_{k\in \D(h)} \A_N(k)
\]
for a set of divisors $\D(h) \subseteq \D_N$, where 
\begin{equation} \label{eq:A_N(k)}
\A_N(k) = \{ i \in \Z_N \colon (i,N)=k\}.
\end{equation}
\end{lemma}

Here $(i,N)$ is the greatest common divisor of $i$ and $N$. The equivalence relation $m_1 \sim m_2$ if $(m_1,N) = (m_2, N)$ already partitions $\Z_N\setminus \{0\}$ into the disjoint union
\[
\Z_N\setminus \{0\} = \bigcup_{k \in \D_N} \A_N(k).
\]
We also have
\[
\A_N(k) = k(\Z_{N/k})^\times,
\]
where $(\Z_{N/k})^\times$ is the multiplicative group of units in the ring $\Z_{N/k}$, so $\A_N(k)$ is $k$ times the elements in $\Z_N$ that are coprime to $k$. In brief, the lemma says that the zero-set of an idempotent is essentially the disjoint union of multiplicative groups.  

See \cite{coven1999tiling}, for example, for a version of this result. The proof in the prime power case is elementary enough, and is reproduced in Section \ref{section:hnull-proof}.

We note one quick corollary.

\begin{corollary}
If $N$ is prime then either $\ZZ(h) = \emptyset$ or $\ZZ(h) = (\Z_N)^\times$. In the latter case $\BB^J = \mathbb{C}^N$.
\end{corollary}

\begin{IEEEproof}
If there is a $k\in \D_N \cap \ZZ(h)$ then  we must have $k=1$ and $\ZZ(h) = (\Z_N)^\times$. In this case $h(m) = \dl_{m0}$ and $\F h=(1,1,\dots,1)$, so $\JJ=[0:N-1]$.
\end{IEEEproof}

We refer to $\D(h)$, which we now know to be $\D_N \cap \ZZ(h)$, as the \emph{zero-set divisors} of $h$. It is helpful to describe $\ZZ(h)$ all at once as 
\begin{equation} \label{eq:zero-set-alternate}
\ZZ(h) = \{i \colon (i,N) \in \D(h)\},
\end{equation}
and then to restate Lemma \ref{lem:oib} as saying that $\II$ is an orthogonal sampling set for $\BB^\JJ$ if and only if 
\begin{equation} \label{eq:zero-set-alternate-2}
(i-j,N) \in \D(h), \quad i \ne j \in \II.
\end{equation}


\subsection{A converse to Lemma \ref{lem:h-null}?} \label{section:h-null-converse}
To study orthogonal sampling sets we will need both Lemma \ref{lem:h-null} and a converse. The converse  would ask to find an idempotent whose zero-set is a prescribed disjoint union of multiplicative groups, and this cannot be done in all cases. For example, let $N=6$ and $\ZZ =\{2,3,4\}$. The set $\ZZ$ can be presented in the form given in Lemma \ref{lem:h-null}, namely $\ZZ=\{2,4\} \cup \{3\}$, but an exhaustive search shows that there is no idempotent $h$ on $\Z_6$ with $\ZZ(h) = \ZZ$.  

There are some cases for which we can easily settle the existence of a converse. For example, when $N=p^M$ is a prime power, the converse to Lemma \ref{lem:h-null} is known to be true. Given a divisor set $\D = \{p^{k_1}, p^{k_2}, \ldots, \}$, the index set $\JJ$ with zero set divisors $\D$ can be constructed as (\cite{Aditya:thesis})
\[
\JJ = \bigcup_{a_0, a_1, \ldots \in [0:p-1]} \left\{\sum_{i}a_ip^{M-k_i-1} \right\}.
\]
We note that the size of $\JJ$ constructed above is $p^{|\D|}$. This is the general form any $\JJ$ that has an orthogonal sampling set:







\begin{theorem} \label{thm:OIS} 
	Let $N=p^M$.  
	Then $\BB^\JJ$ has an orthogonal sampling set if and only if $|\JJ|=p^{|\D(h)|}$.
\end{theorem}

Theorem \ref{thm:OIS} appears in (and can be deduced easily from) many results in literature, for example  \cite[Theorem~B2]{coven1999tiling}, though it is equivalent to Corollary \ref{thm:max-clique-size} in the next section.

 \section{Difference Graphs and Maximal Cliques} 
 
 \label{section:difference-graphs}

 When $N$ is a prime power Theorem \ref{thm:OIS} gives a complete answer to the question of when a space $\BB^\JJ$  has an orthogonal sampling set. We can formulate the problem  in the language of graph theory, and this has significant consequences because of the special structure of the graphs involved.

Let $\G(h)$ be the graph with vertices from $\Z_N$, and with an edge between two vertices $i_1,i_2\in \Z_N$ if $h(i_1-i_2)=0$. We call this the \emph{difference graph} of $h$. We will also denote the graph so constructed as $\G_N(\D(h))$ or simply $\G_N(\D)$, wherever appropriate. At this point we want to note that though such graphs can be constructed from any set of divisors $\D$, it is not clear if the set of divisors $\D$ comes from an idempotent. A converse of Lemma \ref{lem:h-null} is required to make such a claim.

First, let make some comments on the structure of difference graphs. These graphs are Cayley graphs \cite{godsil2013algebraic} defined on a cyclic group, and hence circulant, which itself is enough to characterize some connectivity properties of the graphs (difference graphs are regular, for example). There is some work in literature on the structure of such graphs. Since, according to Lemma \ref{lem:h-null}, the zero-set can be written
$\ZZ(h) = \{i \colon (i,N) \in \D(h)\}$, as in \eqref{eq:zero-set-alternate}, our difference graphs are also what have been called GCD-graphs, see \cite{bavsic2009clique}. Note that GCD-graphs can be constructed starting from \emph{any} divisor set $\D$, but for difference graphs these divisor sets must come from an idempotent. Thus difference graphs are a subclass of GCD-graphs: see \cite{bavsic2009clique}, \cite{ilic2010chromatic} for some results on the clique and chromatic numbers of GCD-graphs. In this paper we are primarily interested in the structure of maximum cliques and perfectness of these graphs, 
and in this section we will show that in several cases $\G(h)$ is a perfect graph. Figures \ref{fig:diff_graph_8_1} and \ref{fig:diff_graph_27_1_9}  are two pictures of difference graphs generated with Mathematica.


\begin{figure}[htb]
 \centering
\includegraphics[scale=0.7,keepaspectratio=true]{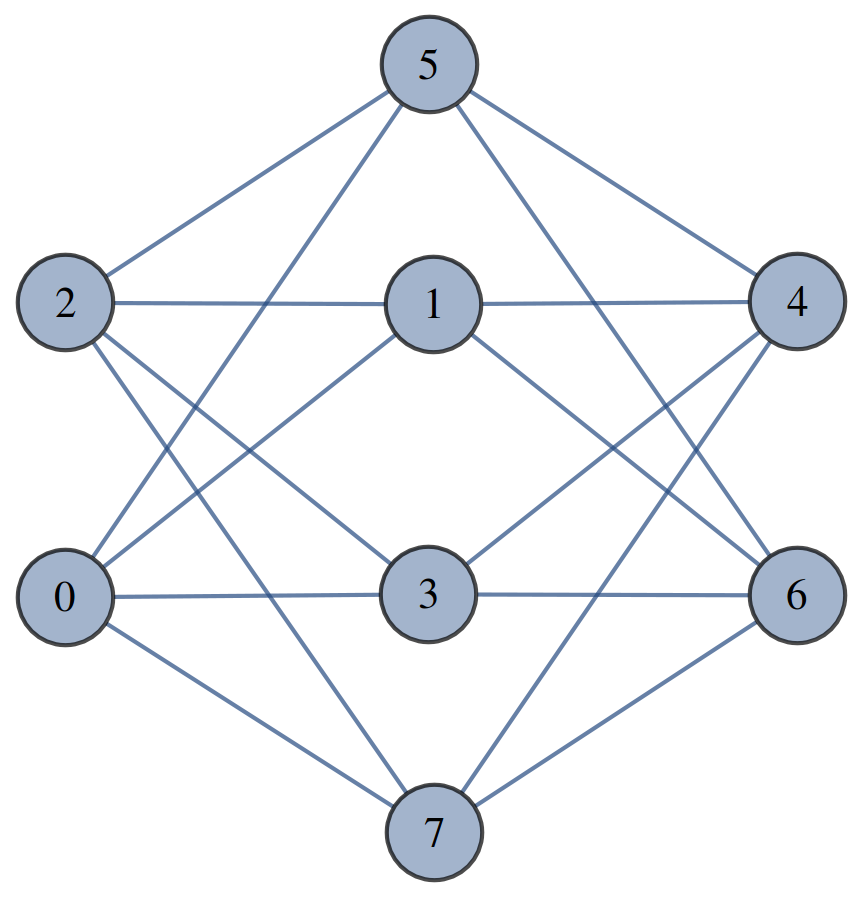}
 
 \caption{Difference graph for $N= 8$, $\D(h) = \{1\}$.}
 \label{fig:diff_graph_8_1}
\end{figure}

\begin{figure}[htb]
 \centering
 \includegraphics[scale=1,keepaspectratio=true]{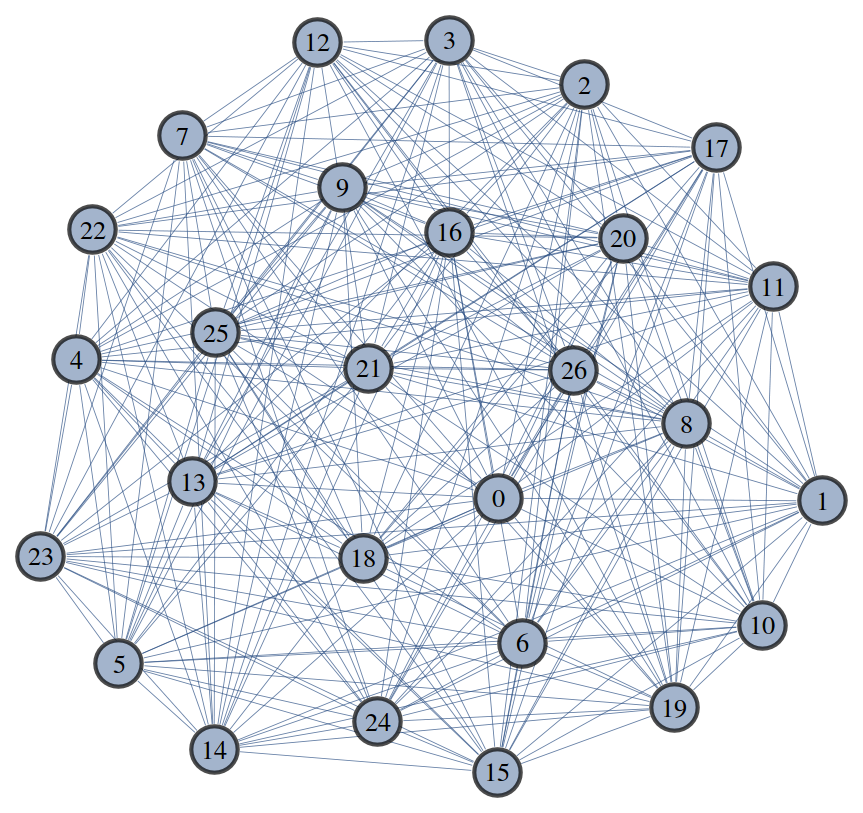}
 
 \caption{Difference graph for $N= 27$, $\D(h) = \{1, 9\}$.}
 \label{fig:diff_graph_27_1_9}
\end{figure}

The key observation relating orthogonal sampling sets to difference graphs comes from Lemma \ref{lem:oib}, which we restate as
\begin{lemma} \label{lemma:diff-graph-cliques}
An index set $\II\subseteq \Z_N$ of size $|\JJ|$ is an orthogonal sampling set for $\BB^\JJ$ if and only if $\II$ determines a maximum clique in $\G(h)$. 
\end{lemma} 
\begin{IEEEproof} Any orthogonal sampling set determines a clique in $\G(h)$. Conversely, any clique whose size is equal to the dimension of $\BB^\JJ$ determines an orthogonal sampling set. The clique determined by an orthogonal sampling set $\II$ must be a maximum clique because it determines an orthogonal basis of $\BB^\JJ$. That is, if a clique (sampling set)  $\II$ can be made larger by adding a vertex $v$ then, by definition of $\G(h)$, the set $\II'=\II\cup\{v\}$ also satisfies
\[
h(i_1-i_2) =0, \quad i_1\ne i_2 \in \II'.
\]
This means that $\II'$ is an orthogonal sampling set for $\BB^\JJ$ which is a contradiction since $|\II'| > |\JJ|$.
\end{IEEEproof}

The following corollary, when combined with Lemma \ref{lemma:diff-graph-cliques}, is essentially a restatement of Theorem \ref{thm:OIS} in terms of cliques. It is  interesting to put it this way because the statement pertains just to idempotents with no references to sampling, orthogonal bases, etc.
\begin{corollary} 
\label{thm:max-clique-size}
Let $N= p^M$ be a prime power and let $h\colon \Z_N \longrightarrow \mathbb{C}^N$ be an idempotent. Then any maximum clique in the difference graph $\G(h)$ is of size $p^{|\D(h)|}$. 
\end{corollary}



\subsection{Perfect difference graphs}  \label{section:perfect-graphs}

Since orthogonal sampling sets correspond to maximum cliques in a difference graph, it is natural to relate the sampling problem to the graph-theoretic (and computational) question of finding maximum cliques. Finding cliques takes exponential time for generic graphs, but in our case the difference graphs have enough structure to solve the problem in polynomial time -- the graphs are perfect when $N$ is a prime power, and in two other cases that we know. Recall that a graph is perfect if for every induced subgraph the chromatic number is equal to the size of a maximal clique. 

\begin{theorem} \label{theorem:perfect-graph} 
Let $h$ be an idempotent and let $\G(h)$ be the associated difference graph. Then $\G(h)$ is perfect when: (a) $N=p^M$; (b) $N=pq$, the product of two primes; 
(c) $N$ and $h$ are such that $|\D(h)| \le 2$ and $|\D(h)^c| \le 2$, where $\D(h)^c = \D_N \setminus \D(h)$.
\end{theorem}

In all cases we prove that $\G(h)$ is a \emph{Berge graph}. The result then follows from the celebrated Strong Perfect Graph Theorem, \cite{perfect-graph}, which states that a graph is perfect if and only if it is a Berge graph. Recall that a graph $\G$ is a Berge graph if every odd cycle with five or more nodes in $\G$ or in $\G^c$ (the complement of $\G$) has a chord. Under each of the assumptions on $N$ the proofs proceed via a series of case distinctions.

\emph{Proof:} Part (a) $N=p^M$: Our starting point is Lemma \ref{lem:h-null}, writing  the zero-set $\ZZ(h)$ as
\begin{align*}
\ZZ(h) = \bigcup_{p^k\in \D(h)} \A_N(p^k),
\end{align*}
for some set of divisors $\D(h) \subseteq \{1,p,p^2,\dots,p^{M-1}\}$.

Suppose that $i_1,i_2,i_3,i_4,i_5, \ldots \in \Z_N$ form a cycle in $\G(h)$.  Then $i_2-i_1 \in \ZZ(h)$, or $(i_2-i_1,N) = p^{k_1}$, say, with $p^{k_1} \in \D(h)$. Similarly $(i_3-i_2,N) = p^{k_2}\in\D(h)$. We can thus write $i_2-i_1=p^{k_1}q_1$ and $i_3-i_2=p^{k_2}q_2$ for some $q_1,q_2$ coprime to $N$. Now consider the following cases.

\textit{Case 1:} ${k_1 < k_2}$. \quad  In this case, $i_3 - i_1 = p^{k_1}(q_1 + q_2p^{k_2 - k_1})$. Since $k_2 - k_1 >0$, we have $p \nmid (q_1 + q_2p^{k_2 - k_1})$. This means that $(i_3-i_1, N) = p^{k_1} \in \D(h)$, and so we have a chord between $i_3$ and $i_1$.

\textit{Case 2:} $k_1 = k_2, \ p \nmid (q_1 + q_2)$.

As in Case 1, we have $i_3 - i_1 = p^{k_1}(q_1 + q_2)$. Since $p \nmid (q_1 + q_2)$ we have $(i_3-i_1, N) = p^{k_1} \in \D(h)$, and so there is a chord between $i_3$ and $i_1$.

\textit{Case 3:} $k_1=k_2, \ p \ | \ (q_1 + q_2)$.

Suppose we have $(q_1 + q_2, N) = p^r, r>0$. Then $i_3 - i_1 = p^{k_1}(q_1 + q_2)$ leads us to $(i_3-i_1, N) = p^{k_1 + r}$, which need not be in $\D(h)$. Hence there need not be an edge between $i_1$ and $i_3$.

Now, $i_4 - i_3 \in \ZZ(h)$, and so  $i_4 - i_3 = p^{k_3}q_3$ for some $q_3$ coprime to $N$. If $k_3 \neq k_2$, then we have a chord between $i_2$ and $i_4$, by the same argument in Case 1 applied to $i_2, i_3, i_4$. This leaves us with the case $k_3 = k_2 = k_1$. Now we have  $i_4 - i_1 = p^{k_1}(q_1 + q_2 + q_3)$. Since $p \ |\ (q_1 + q_2)$ and $q_3$ is coprime to $N$, it follows that $p \nmid (q_1 + q_2 + q_3)$ and so $(i_4 - i_1, N) = p^{k_1} \in \D(h)$. Hence we have an edge between $i_1$ and $i_4$. 

In all cases the cycle has a chord. 

A similar proof holds for cycles in $\G(h)^c$, with $\D(h)$ replaced by $\D(h)^c = \D_N\setminus \D(h)$ which, in this case, is still a set of prime powers.  Thus every odd cycle in $\G(h)$ and in $\G(h)^c$ with at least 5 nodes has a chord and we conclude that $\G(h)$ is a Berge graph, hence perfect.

In fact we have proved that for $N=p^M$ the difference graphs are $P_4-$free or cographs \cite{brandstadt1999graph}.
\smallskip

Part (b),  $N=pq$: Then $\D(h) \subset \D_{pq} = \{1,p,q\}$.  
Take the extreme case, when $\D(h) = \{1,p,q\}$. Then by Lemma \ref{lem:h-null} the zero-set is the set of all nonzero indices, $\ZZ(h) =[1:N-1]$. So all the vertices in the difference graph are connected to each other and every cycle has a chord, while $\G(h)^c = \emptyset$. The remaining cases to show that $\G(h)$ is perfect are then covered by part (c).

Now to part (c) of the theorem. In the following we will take a cycle of size five, but the argument can be generalized to any cycle of odd size. Assume that $i_1, i_2, i_3, i_4, i_5$ form a cycle in $\G(h)$. Let $r_k = (i_k - i_{k+1}, N)$. Since $\D(h)$ is at most of size $2$ there are two cases.

 \emph{Case 1:} $\hZd \neq \{1\}$, i.e. $\hZd$ includes a prime.
 
 In this case all the $r_k$ are divisible by either of at most two primes, say $p$ and $q$. For the cycle not to have a chord between vertices $i_s, i_t$, we need $i_s - i_t$ to be coprime to both $p$ and $q$. Now consider any two consecutive edges, say $i_1-i_2$ and $i_2 - i_3$. Suppose both these edges correspond to divisibility by \emph{the same prime}, i.e., both $r_1$ and $r_2$ are divisible by, say, $p$. It follows that $i_3-i_1$ is divisible by $p$ as well, which means the edge $i_3-i_1$ exists in $\G(h)^c$.  Hence all consecutive edges must correspond to divisibility by different primes, as indicated in Fig. \ref{fig:pq2}. However since the cycle is odd in size, this is impossible.
 \begin{figure}[htb]
\centering
\begin{tikzpicture}[scale=1] 
\tikzstyle{every node}=[font=\small]
\draw [fill] (0,4) circle [radius=0.07] node [below] {$i_1$} -- (2,3) circle [fill, radius=0.07] node [right] {$i_2$} node [midway] {$p$} -- (2, 1) circle [fill, radius=0.07] node [below] {$i_3$} node [midway] {$q$} --  (0, 0) circle [fill, radius=0.07] node [below] {$i_4$} node [midway] {$p$} -- (-1, 2) circle [fill, radius=0.07] node [below] {$i_5$} node [midway] {$q$} -- (0,4) node [midway] {$p$};
\end{tikzpicture}
		\caption{Case 1: Assume $i_1, i_2, i_3, i_4, i_5, \ldots$ form a cycle in $\G(h)$. Each of the edges must correspond to divisibility by either $p$ or $q$.  If the cycle does not have any chords, then consecutive edges must correspond to divisibility by different primes, as indicated. Since the cycle is odd in size, this is impossible to obtain.} 
	\label{fig:pq2}
\end{figure}
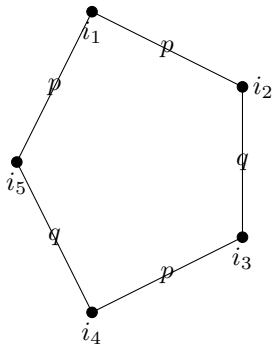

 \emph{Case 2:} $\hZd= \{1\}$.    Since $|\hZd^c| \leq 2 $, we can assume that $\D(h)^c  \subseteq \{p, q\}$.  In this case all the $r_k$ are coprime to $N$. For a chord $i_s - i_t$ not to exist, $i_s - i_t$ must be divisible by either $p$ or $q$, and thus each chord falls into one of these two groups. Now consider two \emph{adjacent} chords , say $i_6 - i_2$ and $i_6 - i_3$, as in Fig. \ref{fig:pq}. Then both $i_6 - i_2$ and $i_6 - i_3$ should be divisible by different primes, for otherwise $i_2 - i_3$ would be divisible by $p$ as well. But this too is impossible according to the following purely geometric observation:

\begin{lemma}
 \label{lem:adj-chords}
 Consider a polygon $C$ with $n$ vertices, where $n$ is odd. We call two chords (diagonals) of $C$ \emph{adjacent} if they form a triangle with one of the sides of $C$ (For example as in  Fig. \ref{fig:pq}). Then it is impossible to divide the chords into two groups such that adjacent chords belong to different groups.
\end{lemma}

\begin{IEEEproof}
Once again we make some case distinctions. 

 \emph{Case 1: $n = 5$}  Suppose the chords in each group are represented by drawing them \emph{dashed} and \emph{dotted}, respectively. Assuming the chord $i_4 - i_2$ to be dotted, we can dash/dot the remaining chords (see Fig. \ref{fig:pqproof1}). We end up with chord $i_1 - i_4$ which has to be both dashed and dotted, a contradiction.

\begin{figure}[htb]
\centering
\begin{tikzpicture}[scale=1]
\tikzstyle{every node}=[font=\small]
\draw [fill] (0,6) circle [radius=0.07] node [below] {$i_1$} -- (1,5) circle [fill, radius=0.07] node [right] {$i_2$} -- (1.25, 4) circle [fill, radius=0.07] node [below] {$i_3$} --  (1, 3) circle [fill, radius=0.07] node [below] {$i_4$} -- (0, 2) circle [fill, radius=0.07] node [below] {$i_5$} -- (-1,3)circle [fill, radius=0.07] node [below] {$i_6$} -- (-1,5)circle [fill, radius=0.07] node [below left] {$i_7$} -- (0,6);
\draw [dashed] (-1,3) -- (1,5) node [midway] {$p$};
\draw [dashed] (-1,3) -- (1.25,4) node [midway] {$q$};

\end{tikzpicture}

		\caption{Case 1 (b): Assume $i_1, i_2, i_3, i_4, i_5, \ldots$ form a cycle in $\G(h)$. } 
			\label{fig:pq}
\end{figure}
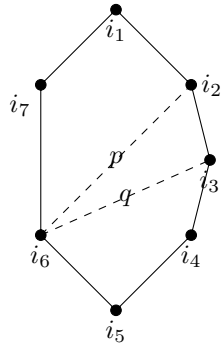

\emph{Case 2: $n \geq 7$}  The argument is similar to the previous case. Assume $i_6 - i_4$ is dotted. Then we can mark the remaining chords as dashed and dotted, as in Fig. \ref{fig:pqproof2}. There is no way to mark the chord $i_1 - i_4$ consistently.
\end{IEEEproof}

\begin{figure}[htb] 
\centering
\begin{tikzpicture}[yscale=1.5]
\tikzstyle{every node}=[font=\small]
\node(v1) at (0,2){};
\node(v2) at (2,1) {};
\node(v3) at (1,0) {};
\node(v4) at (-1,0) {};
\node(v5) at (-2,1) {};
\draw[fill] (v1) circle [radius=0.07] node [above] {$i_1$}  -- (v2)circle [radius=0.07]node [right] {$i_2$}  -- (v3)circle [radius=0.07] node [right] {$i_3$}  -- (v4)circle [radius=0.07] node [below right] {$i_4$}   -- (v5)circle [radius=0.07] node [below] {$i_5$}  -- (v1);
\draw [dotted] (v4) -- (v2);
\draw [dashed] (v2) -- (v5);
\draw [dotted] (v5) -- (v3);
\draw [dashed] (v1) -- (v3);
\draw [dashdotted, very thick] (v1) -- (v4);
\end{tikzpicture}
		\caption{Assume the chord $i_4 - i_2$ is dotted. Then the chord $i_1 - i_4$ has to be both dashed and dotted, a contradiction.} 	\label{fig:pqproof1}
\end{figure}
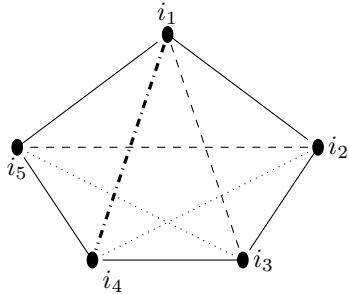

\begin{figure}[htb]
\centering
\begin{tikzpicture}[scale=1]
\tikzstyle{every node}=[font=\small]
\node(v1) at (0,6){};
\node(v2) at (2,5) {};
\node(v3) at (2.75,3) {};
\node(v4) at (2.5,1) {};
\node(v5) at (0,0) {};
\node(v6) at (-2,1) {};
\node(v7) at (-2,3.5) {};
\draw[fill] (v1) circle [radius=0.07] node [above] {$i_1$}  -- (v2)circle [radius=0.07]node [right] {$i_2$}  -- (v3)circle [radius=0.07] node [right] {$i_3$}  -- (v4)circle [radius=0.07] node [below right] {$i_4$}   -- (v5)circle [radius=0.07] node [below] {$i_5$}  -- (v6) circle [radius=0.07] node [below left] {$i_6$} -- (v7)circle [radius=0.07] node [left] {$i_7$}   -- (v1);
\draw [dotted] (v6) -- (v4); 
\draw [dotted] (v6) -- (v2);
\draw [dotted] (v2) -- (v4);
\draw [dashed] (v6) -- (v3); 
\draw [dashed] (v6) -- (v1);
\draw [dashed] (v1) -- (v3);
\draw[dashdotted, very thick] (v1) -- (v4);
\end{tikzpicture}
		\caption{Assume $i_6 - i_4$ is dotted. Then we can mark all the chords shown as dashed/dotted, except there is no way to mark $i_1 - i_4$ consistently. } 
			\label{fig:pqproof2}
\end{figure}
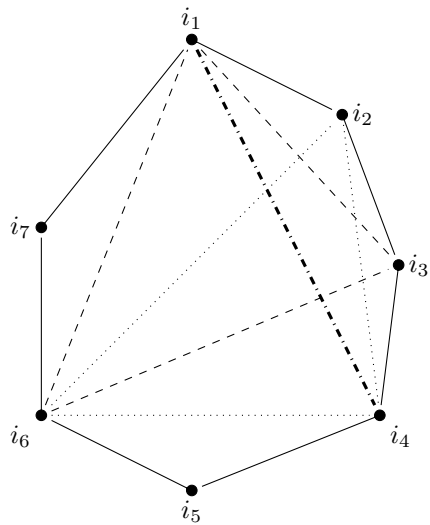

With Lemma \ref{lem:adj-chords} the final case of Theorem \ref{theorem:perfect-graph} is settled. \qed 
 \begin{figure}[htb]
 	\centering
 	\begin{tikzpicture}[scale=0.75]
 	\tikzstyle{every node}=[font=\small]
 	\node(v1) at (-1,6){};
 	\node(v2) at (2,6) {};
 	\node(v3) at (4,3) {};
 	\node(v4) at (0,0) {};
 	\node(v5) at (-2,3) {};
 	\draw[fill] (v1) circle [radius=0.07] node [above left] {$1$}  -- (v2)circle [radius=0.07]node [right] {$4$}  -- (v3)circle [radius=0.07] node [below right] {$3$}  -- (v4)circle [radius=0.07] node [below left] {$31$}   -- (v5)circle [radius=0.07] node [left] {$12$}  -- (v1);
 	\end{tikzpicture}
 	\caption{$N=72$. For the difference graph $\G_{72}(\{1,3,4,12\})$, the cycle formed by the nodes $1, 4, 3, 31, 12$ is shown. It has no chords. } 
 	\label{fig:72-not-perfect}
 \end{figure}
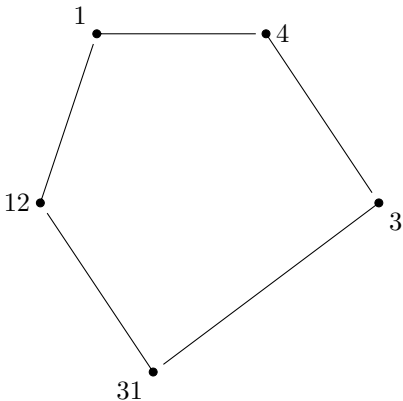

\section{Further comments for $N=p^M$}
In the case when $N=p^M$, Theorem \ref{theorem:perfect-graph} establishes that any difference graph is perfect, and maximum cliques can be determined in polynomial time. In fact, for $N=p^M$ we can say more: all difference graphs (and their complements) are well covered \cite{plummer1993well}. In other words, any maximal clique is maximum, and so maximum cliques can be found by a simple greedy algorithm.

\begin{theorem}
	\label{thm:maximal-maximum}
	When $N=p^M$, all maximal cliques in the difference graph $\G(\D)$ are of size $p^{|\D|}$. 
\end{theorem} 
\begin{IEEEproof}
We prove this by induction on $|\D|$. Suppose $p^k$ is the largest divisor in $\D$. For any maximal clique $\chi$ in $\G(\D)$, we note that $\chi' = \chi \mod p^k$ is a maximal clique in $\G(\D\setminus\{p^k\})$ and so
\[
|\chi'| = p^{|\D| - 1}.
\] 
Any element of $\chi$ is of the form $\alpha p^k + \beta$ for $\beta  \in \chi'$, and for two elements $\alpha_1 p^k + \beta_1$ and  $\alpha_2 p^k + \beta_2$ of $\chi$ with $\alpha_1 \neq \alpha_2$, $\alpha_1 - \alpha_2$ must be coprime to $p$. Thus it follows that for $\chi$ to be maximal, $\alpha \text{ mod }p$ takes on all the $p$ values $\{0,1,2,\ldots,p-1 \}$, and so 
\[
|\chi| = p|\chi'| = p^{|\D|}.
\]

For the case when $\D = \emptyset$, the size of any maximal clique is $1$, thus completing the proof. 
\end{IEEEproof}
The theorem is not true for arbitrary $N$. As an example, consider $N=18$ and $\D = \{2,3\}$. In the difference graph $\G_{18}(\{2,3\})$, we note that 
$\{0,3\}$ is maximal, and so is $\{0,2,4\}$.

\subsection{Counting the number of orthogonal sampling sets} \label{section:counting}

Lemma \ref{lemma:diff-graph-cliques} allows us to count the number of orthogonal sampling sets. We first find the number of orthogonal sampling sets for a given $\BB^\JJ$ with $|\JJ|=d$, for this  let $\n_N(\D)$ denote the number of maximal cliques in $\G_N(\D)$ for a set of divisors $\D$ of $N$.

\begin{lemma}
	\label{lem:oib-recursion}
	Suppose $N = p^M$ is a prime power. Let $\D$ be a non empty set of divisors of $N$, and let $p^{l}$ be the smallest element of $\D$. Then, the number maximal cliques $\n_N(\D)$ is given by
\[
 \n_N(\D) = p^{l}\left(\n_{N_1}(\D_1) \right)^p,
\]
where
\[
\D_1 = \frac{1}{p^{l+1}}\left(\D \setminus \{p^l\}\right),  \quad N_1 =\frac{N}{p^{l+1}}.
\] 
If $\D$ is empty, we have $\n_N(\emptyset) = N = p^M.$ 
\end{lemma}

\begin{IEEEproof}
	Note that if $\D$ is empty, every vertex is a max clique (max clique size is $1$ by Theorem \ref{thm:max-clique-size}) and so the number of max cliques is simply the number of vertices in the graph, i.e. $N$.
	
	Now assume $\D$ is non empty. First we establish the result assuming $p^l=1$.	For any maximal clique $\chi$ of $\G_N(\D)$, consider the congruence classes modulo $p$:
	\[
	\chi_k = \{i \in \chi: i \mod p = k\}, \quad  0 \leq k < p.
	\]
	Note that for any $i \in \chi_k$, 
	\begin{itemize}
		\item $p \ | \ i-k$, so $(i-k,N) = p\ (\ (i-k)/p, \ N/p)$, and
		\item $(i-k,N) \in \D \setminus \{1\}$. 
	\end{itemize}
	Combining these two, we note that 
	\[
	\frac{\chi_k - k}{p} \text{ is a maximal clique in }\G_{N/p}\left(\frac{\D \setminus \{1\}}{p}\right).
	\] 
	Thus we see that any clique in $\G_N(\D)$  is composed of $p$ cliques from $\G_{N_1}(\D_1)$. Each of these cliques is multiplied by $p$, and `raised' to a different congruence class modulo $p$, and taking the union of these `raised' cliques gives us a clique in $\G_N(\D)$. We note that this correspondence is one-to-one, and so
	\[
	 \n_N(\D) = \left(\n_{N_1}(\D_1) \right)^p, \quad N_1 = N/p,
	\]
	thus completing the proof.
	
	Now consider the case for an arbitrary smallest divisor $p^l$ (not necessarily $1$). For any two elements $i_1,i_2$ of a maximal clique $\chi$, we see that $p^l \ | \ i_1 - i_2$, so that 
\[
i_1\mod p^l = i_2 \mod p^l = c \text{ (say) }.
\]

It follows that all elements of $\chi - c$ are divisible by $p^{l}$. At this point we note that the number of possible ways to pick $c$ given $\chi - c$ is simply, $p^l$: the number of congruence classes modulo $p^l$.

 Now note that for any $i \in \chi$, 
\[
(i-c,N) = p^{l} ((i-c)/p^l,N/p^l),
\]
so that $((i-c)/p^l, N) \in \D/p^l$. So if we consider $\chi' = (\chi - c)/p^l$, we see that $\chi'$ is a clique in $\G_{N/p^l}(\D/p^l)$. Since $1 \in \D/p^l$, the number of ways to construct $\chi'$, from the previous paragraph, is 
\[
\left(\n_{N_1}(\D_1) \right)^p, \quad N_1 = N/p^{l+1}, \quad \D_1= \left(\frac{\left(\D/p^l \right)\setminus \{1\}}{p}\right) = \frac{1}{p^{l+1}}\left(\D \setminus \{p^l\}\right).
\]
Since the number of ways to construct the translate $c$ is $p^l$, the theorem follows.

\end{IEEEproof}

We next focus on finding the total number of maximal cliques of size $d = p^{|\D|}$ in $\G_N(\D)$. Suppose $\D = \{ p^{l_0}, p^{l_1}, p^{l_2},\ldots p^{l_{\log d-1}}\}$ with $l_0<l_1 <l_2 < \ldots < l_{\log d -1 }$. 
To expand upon the recursion from Lemma \ref{lem:oib-recursion}, first note that the smallest element of $\D_1$ is $p^{l_1-l_0-1}$. We can define
\[
N_2 = N/{p^{l_1+1}}, \D_2 = \frac{1}{p^{l_1+1}}\left(\D \setminus \{p^{l_0}, p^{l_1}\}\right)
\]
to continue the recursion. We see that
\begin{align*}
\n_N(\D) = p^{l}\left(\n_{N_1}(\D_1) \right)^p \\
&= p^{l_0}p^{(l_1 - l_0 - 1)p}\left( \n_{N_2}(\D_2)  \right)^{p^2} \\
&= \cdots  
= p^{\sum_{i=0}^{\log d -1 } (l_{i+1} - l_i - 1)p^i},
\end{align*}
where we take $l_{\log d} = M$. Now we also take $l_{-1} = -1$, and let 
\[
r_i = l_{i} - l_{i-1} - 1.
\]
Then $r_i$ represents the `gap' between successive divisors in $\D$. Note that
\[
\sum_{i=0}^{\log d} r_i = \sum_{i=0}^{\log d} (l_{i} - l_{i-1} - 1) = M - \log d.
\]

So we have

\begin{lemma}
	\label{lem:oib-count-given-j}
Suppose $N=p^M$ is a prime power. The number of orthogonal sampling sets for $\BB^\JJ \subseteq \mathbb{C}^N$ is given by $\n_N(\D) = p^{\lambda(\mathbf{r})}$ where 
\[
\lambda(\mathbf{r}) = \sum_{i=0}^{\log d}r_i p^i, \quad \mathbf{r} = (r_0,r_1, r_2, \dots),
\]
where $\D = \{ p^{l_0}, p^{l_1}, p^{l_2},\ldots p^{l_{\log d-1}}\}$ with $l_0<l_1 <l_2 < \ldots < l_{\log d -1 }$, and $r_i =  l_{i+1} - l_i - 1$.
\end{lemma}

We obtain the total number of orthogonal sampling sets of size $d$ by summing over all possible $(r_0, r_1, \ldots, r_{\log d})$ that sum to $ M - \log d$:
\begin{align*}
&\text{Number of orthogonal sampling sets of size }d \\
&\hspace{.25in}= \sum_{\sum r_i = M - \log d} p^{\lambda(\mathbf{r})} \\
& \hspace{,25in}= \sum_{\sum r_i = M - \log d}p^{ \sum_{i=0}^{\log d}r_i p^i.}
\end{align*}
From the form of this expression we observe that the count is given by a $(\log d)$-fold convolution:
\[
\begin{aligned}
&\text{Number of orthogonal sampling sets of size }d\\
&\quad   = \left(f_0 * f_1 *f_2*\ldots*f_{\log d} \right) (M - \log d),
\end{aligned}
\]

where $f_i: \mathbb{Z} \longrightarrow \mathbb{Z} $ is 
\[
f_i(r) = \begin{cases} p^{rp^i} \quad r\geq 0  \\ 
0 \quad \text{otherwise}.
\end{cases}
\]

The generating function of $f_i$ is 
\[
F_i(x) = \sum_r x^r f_i(r) = \sum_{r=0}^{\infty}p^{rp^i}x^r = \frac{1}{1- xp^{p^i}},
\]
so the generating function for $f_0 * f_1 *f_2*\ldots*f_{\log d} $ is the product of individual generating functions. In conclusion:
\begin{theorem}
\label{thm:oib-total-count}
For prime powers $N = p^M$ and $d = p^{\log d}$, define the generating function
\begin{equation}
\Theta_{d}(x) = \prod_{i=0}^{\log d}\frac{1}{1-xp^{p^i}}.
\end{equation}
The number of orthogonal sampling sets of size $d$  in $\Z_N$ is the coefficient of $x^{M - \log d}$ in $\Theta_{d}(x)$. \qed
\end{theorem}

To get an idea of the numbers, we plot
\begin{align*}
\theta_N(d) =
\frac{1}{d}\log (\text{Number of orth. sampling sets of size } d).
\end{align*}
Figures \ref{fig:oibcount_2_10}  and \ref{fig:oibcount_3_20} plot $\theta_N(d)$ vs $\log d$ for various prime powers $N$. The function $\theta_N(d)$ is \emph{not} equal to the linear function that jumps out in the plots, but in the examples we have tried it appears to be remarkably close.
\begin{figure}[htb]

\includegraphics[scale = 0.6]{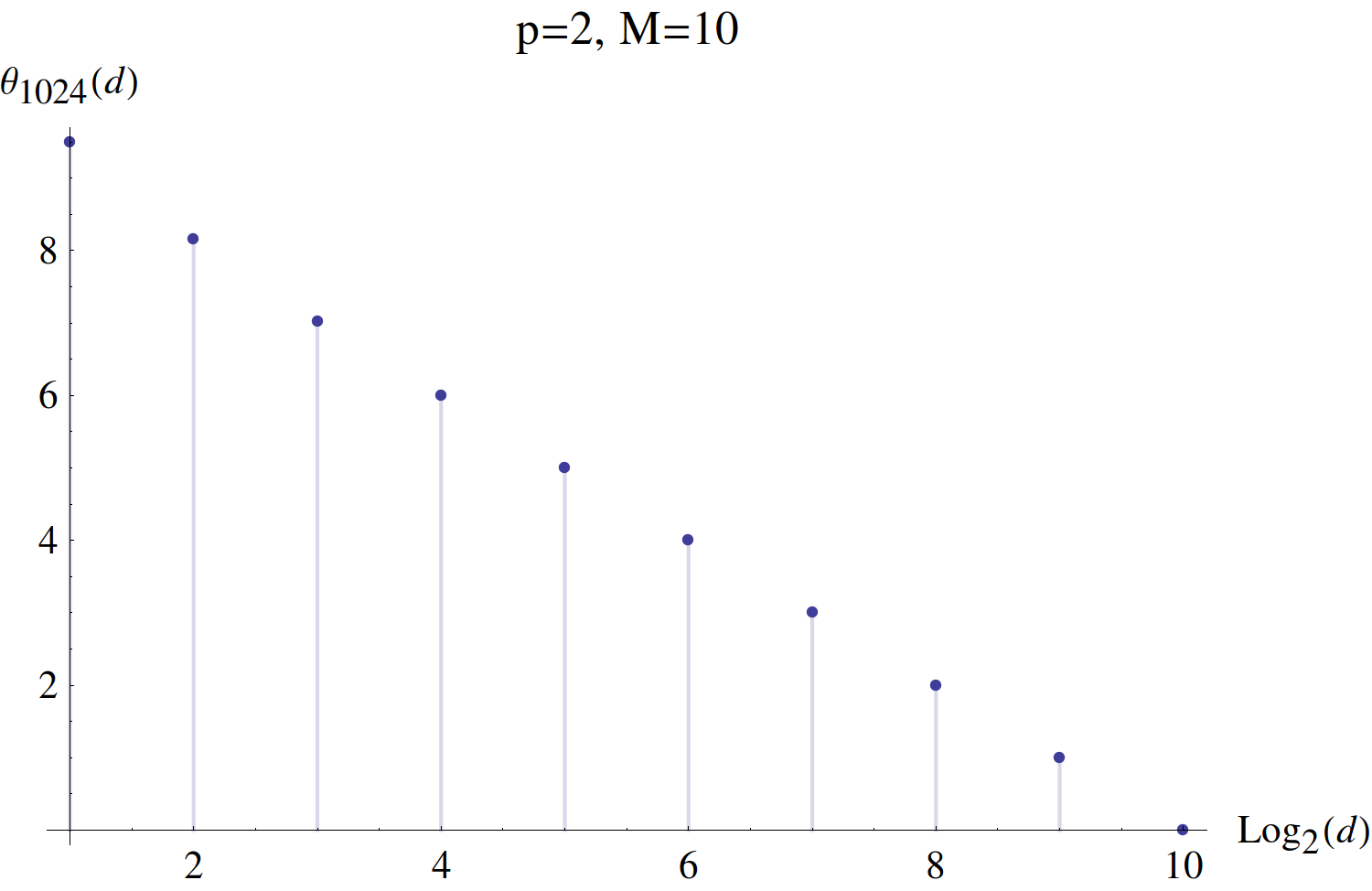}
\caption{}
\label{fig:oibcount_2_10}
\end{figure}

\begin{figure}[htb]

\includegraphics[scale = 0.7]{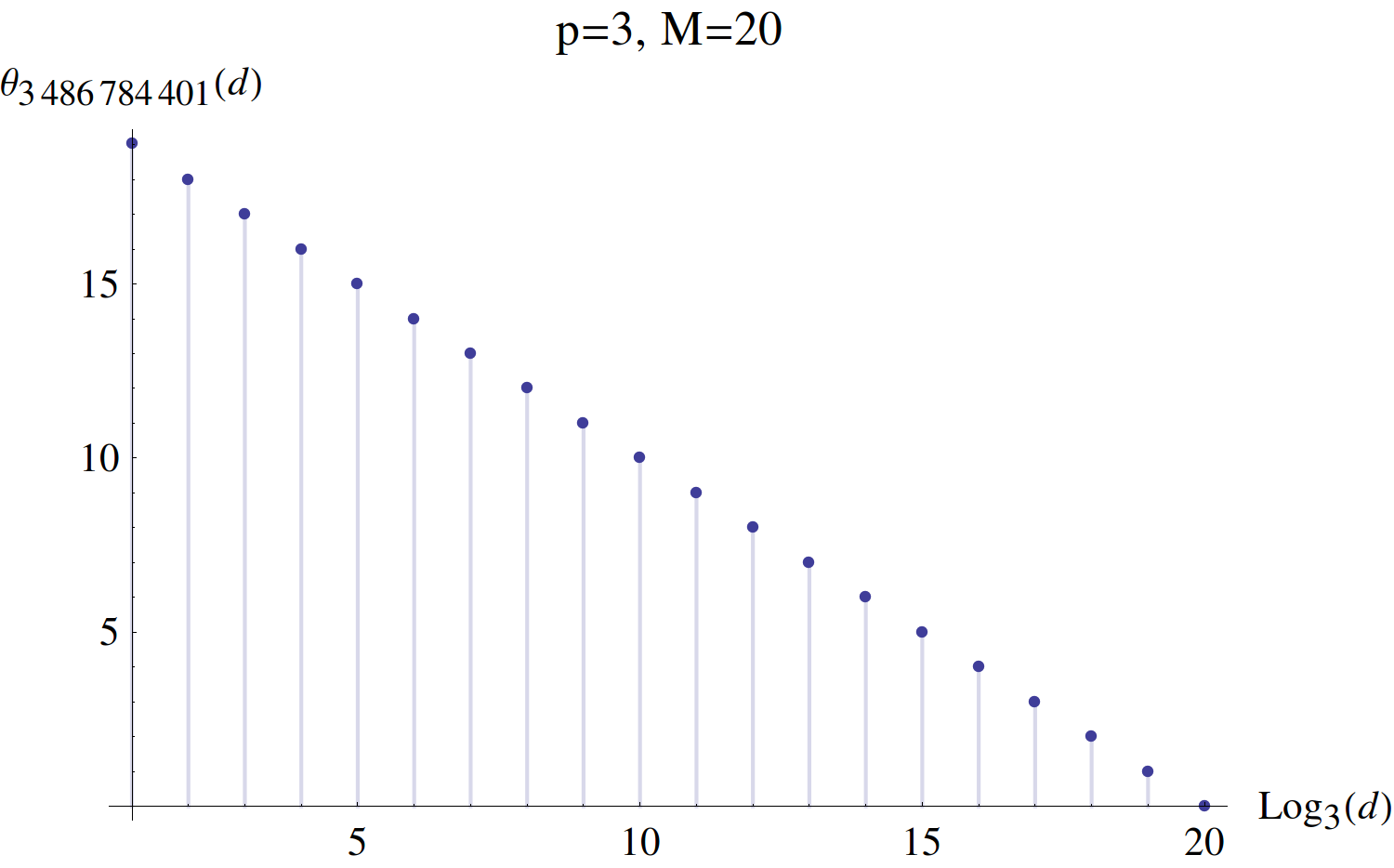}
\caption{}
\label{fig:oibcount_3_20}
\end{figure}

For an  example of Theorem \ref{thm:oib-total-count}, when $d=4$ and $p=2$ we have
\begin{align*}
\Theta_4(x) &= \prod_{i=0}^{2}\frac{1}{1-x2^{2^i}} \\
&=  \frac{1}{(1-x2^1)(1-x2^2)(1-x2^4)}. 
\end{align*}
The number of orthogonal sampling sets $\II$ of size $4$ in $\Z_8$ is the coefficient of $x^1$, which we find to be $22$.  This is out of a total of $70$ index sets of size $4$. The number of orthogonal sampling sets of size $4$ in $\Z_{16}$ is the coefficient of $x^2$, which is  $380$, and is out of $1820$. 

\section{Perfectness for arbitrary $N$}
\label{section:perfectness-abritrary-N}
The case of $N$ having more than one prime factor assumes significance when dealing with higher dimensional orthogonal interpolation. For example, in the typical two dimensional setting, suppose we assume each of the dimensions to be $p_1^{M_1}$ and $p_2^{M_2}$ respectively ($p_1 \neq p_2$ are primes). The resulting two dimensional orthogonal interpolation problem can be easily seen to be equivalent to a one dimensional orthogonal interpolation problem with $N=p_1^{M_1}p_2^{M_2}$ \cite{Aditya:thesis}. Investigating the relationship between existence of orthogonal sampling sets and properties of the corresponding difference graph for an arbitrary dimension is an intriguing question, which we attempt to address in this section. One naturally interesting question is to what extent perfectness, as in Theorem \ref{theorem:perfect-graph}, holds.

 Examples disproving perfectness and other properties for GCD-graphs \cite{bavsic2009clique}, \cite{ilic2010chromatic} cannot directly apply to difference graphs. Here is an example. Let $N= 8 \times 9$ and let $\ZZ=\A_N(1)\cup \A_N(3)\cup\A_N(4)\cup\A_N(12)$ (all elements of $\Z_{N}$ whose greatest common divisor with $N$ is $1,3,4$ or $12$). We take $\G$ to be the GCD-graph determined by $\ZZ$, i.e., there is an edge between $i_k$ and $i_\ell$ if $i_k-i_\ell \in \ZZ$. Now consider the nodes $1,4,3,31,12$. Figure \ref{fig:72-not-perfect} shows the cycle on these nodes, and $\G$ is not perfect (this example is due to Sivatheja Molakala). \emph{But we do not know if $\ZZ$ is the zero-set of an idempotent.} It is computationally infeasible to check this, and we do not know to what extent the converse of Lemma \ref{lem:h-null} holds when $N$ is not a prime power.

 As a special case, when $\D(h)$ is a singleton the graph $\G(h)$ becomes a unitary Cayley graph, \cite{klotz2007some}, and such graphs are shown to be perfect when either $N$ is even or $N$ is odd with at most two prime factors. When $\D(h)$ is a singleton, a converse to Lemma \ref{lem:h-null} can easily be seen to be true \cite{Aditya:thesis}, and thus the result of \cite{klotz2007some} generalizes Theorem \ref{theorem:perfect-graph} to arbitrary $N$ when $\D$ is a singleton. For $\D$ of arbitrary size; perfectness of the difference graph $\G(\D)$ has not been investigated, nor has the existence of an idempotent that generates $\D$.  Thus, in addition to investigating GCD-graphs thoroughly, a more encompassing converse to Lemma \ref{lem:h-null} is crucial for investigating the higher dimensional interpolation problem.






The next natural question is to ask if we can impose specific restrictions on $\JJ$ such that the corresponding difference graph is perfect. For reasons to be elaborated on in Section \ref{sec:other-graph-parameters}, one reasonable restriction we could think of is that $\JJ$ should \emph{tile} $\Z_N$. Take for instance $\JJ = \{0,6,9,15\}$, it can be verified that $\JJ$ satisfies the following properties of interest:

\begin{enumerate}
	\item The index set $\JJ$ \emph{tiles} $\Z_{36}$ with $\KK= \{0,4,8,12,16,20,24,28,32\}$, in other words the sumset $\JJ\oplus \KK = \{j+k \mod 36, j\in \JJ, k \in \KK\}$ is equal to $\Z_{36}$, and
	\item The bandlimited space $\BB^\JJ$ has an orthogonal sampling set.
\end{enumerate}
Finding $h =\F^{-1}1_\JJ$, we see that $\D_\JJ = \{2,3,6,9,18\}$. Consider the induced cycle shown in Figure \ref{fig:tiling-perfectness}, which proves that $\G(h)$ is not perfect. This suggests that perfectness is not likely the correct property to investigate for a general $N$. 




\begin{figure}[htb]
	\centering
	\begin{tikzpicture}[scale=1]
	\tikzstyle{every node}=[font=\small]
	\node(v1) at (0,6){};
	\node(v2) at (2,4) {};
	\node(v3) at (2,2) {};
	\node(v4) at (-2,2) {};
	\node(v5) at (-2,4) {};
	\draw[fill] (v1) circle [radius=0.07] node [above] {$0$}  -- (v2)circle [radius=0.07]node [right] {$2$}  -- (v3)circle [radius=0.07] node [below right] {$4$}  -- (v4)circle [radius=0.07] node [below left] {$7$}   -- (v5)circle [radius=0.07] node [left] {$9$}  -- (v1);
	\end{tikzpicture}
	\caption{A length $5$ cycle in $\G_{36}(\{2,3,6,9,18\})$ with no chord. This difference graph corresponds to $\JJ = \{0,6,9,15 \}$, which tiles $\Z_{36}$. } 
	\label{fig:tiling-perfectness}
\end{figure}
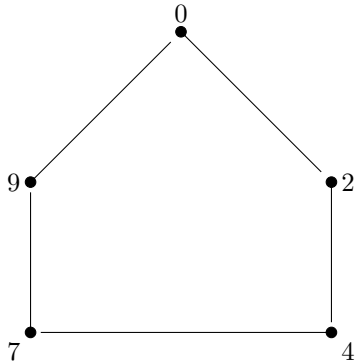

\section{Clique, chromatic and Lov{\'a}sz numbers}

\label{sec:other-graph-parameters}
In this section, we attempt to investigate the clique, chromatic numbers (their gap in particular), and their implications for the underlying bandlimited space $\BB^\JJ$. Denote the size of the largest clique in $\G$ by $\mu(\G)$, and the chromatic number by $\chi(\G)$. Recall that $\mu(\G) \leq \chi(\G)$, and the chromatic gap $\chi(\G) - \mu(\G)$ is zero if the difference graph is perfect.

Also recall that $\G^c$ is the graph on the same vertices as $\G$ formed with the edges that are missing from $\G$. In our case, we see that for a difference graph $\G$, $\G^c$ corresponds to the difference graph constructed from the divisors that are missing from $\G$, i.e.
\[
\G^c(\D) = \G(\D^c), \quad \D^c = \D_N \setminus \D.
\]
Next, the Lov{\'a}sz number (\cite{lovasz1979shannon}) is 
\[
\vartheta(\G^c) = \min_{u, V} \max_{m}\frac{1}{\left| \langle u, v_m \rangle \right|^2}\,,
\]
where $u$ is a unit vector and $V= \{v_0, v_1, \ldots, v_{N-1}\}$ is an orthonormal representation of $\G^c$, i.e. 
\[
\langle v_m,v_n \rangle = 0 \text{ whenever }m,n \text{ are connected in } \G.
\]

\noindent Here $\langle . \rangle$ denotes the standard inner product.
The Lov{\'a}sz number is sandwiched between the clique number $\mu(\G)$   and the chromatic number $\chi(\G)$ of $\G$:
\[
\mu(\G) \leq \vartheta(\G^c) \leq \chi(\G).
\]

For difference graphs $\G$ constructed from $h = \F^{-1}1_\JJ$, we know from Lemma \ref{lem:oib} that $\mu(\G) \leq |\JJ|$, and in particular if 
 $\BB^\JJ$ has an orthogonal sampling set then $\mu(\G) = |\JJ|$. The following theorem is a stronger version of this observation.

\begin{theorem}
	\label{thm:lovasz-bound}
	Let $\G$ be a difference graph constructed from an idempotent $h = \F^{-1} 1_\JJ$. Then $\vartheta(\G^c) \leq |\JJ|$, where $\vartheta(\G^c)$ is the Lov{\'a}sz number (\cite{lovasz1979shannon}) of the graph $\G^c$. If $\BB^\JJ$ has an orthogonal sampling set, then $\vartheta(\G^c) = |\JJ|$.
\end{theorem}

\begin{IEEEproof}Note that $m,n$ are connected in a difference graph $\G(h)$ only if $h(m-n) = \langle\tau^mh,\tau^nh \rangle=0$. Thus $v_m = \tau^mh/\Vert h\Vert$ forms an orthonormal representation of $\G^c(h)$, and so
\[
	\vartheta(\G^c) =  \min_{u, V} \max_{m}\frac{1}{\left| \langle u, v_m \rangle \right|^2} 
	\leq \min_{u} \max_{m}\frac{\Vert h \Vert^2}{\left| \langle u, \tau^mh \rangle \right|^2}.
\]	
	Fix some $j \in \JJ$, and let $\delta_j$ be the corresponding canonical basis vector. Set $u$ to be the unit vector $\sqrt{N}\F^{-1}\delta_j$. Then since $\langle \sqrt{N}\F^{-1}\delta_j, \tau^mh \rangle = N^{-1}\langle \sqrt{N}\delta_j, \F(\tau^mh) \rangle$, we obtain
	\[
	\vartheta(\G^c) \leq \max_{m}\frac{\Vert h \Vert^2}{\left| \langle \sqrt{N}\F^{-1}\delta_j, \tau^mh \rangle \right|^2} = \max_{m}\frac{N\Vert h \Vert^2}{\left| \langle \delta_j, \F (\tau^mh) \rangle \right|^2} = \max_{m}\frac{N\Vert h \Vert^2}{\left| e^{2\pi i mj/N} \right|^2} =   N\Vert h \Vert^2.
	\]
	
	Since $N\Vert h\Vert^2 = |\JJ|$, it follows that $\vartheta(\G^c) \leq |\JJ|$. 
	Moreover, if $\BB^\JJ$ has an orthogonal sampling set then we have $ |\JJ|=\mu(\G) \leq \vartheta(\G^c) \leq |\JJ|$, and thus $\vartheta(\G^c) = |\JJ|$.
	
	\end{IEEEproof}
An even stronger version of Theorem \ref{thm:lovasz-bound} would require us to investigate if the chromatic number of $\G$ is bounded by $|\JJ|$. While we are unable to verify this, we do have one interesting observation.

\begin{theorem}
	\label{thm:cliques -tiling}
	Suppose $\BB^\JJ$ has an orthogonal sampling set, and $\G$ be the associated difference graph. If the chromatic number of $\G$ is equal to $|\JJ|$ then $\mu(\G)\mu(\G^c) = N$ and
	 any max clique in $\G$ \emph{tiles} $\Z_N$.

\end{theorem}

We say that a set $\II \subseteq\Z_N$ \emph{tiles} $\Z_N$ if there exists a set of translates $\bar{\II}$ such that every $a\in \Z_N$ can be written uniquely as $a=i+\bar{i}$ $\mod N$, with $i \in \II$, $\bar{i} \in \bar{\II}$. 
\begin{IEEEproof}
	Suppose $C$ is a max-clique in $\G$ and $\bar{C}$ is a max-clique in $\G^c$.  If $i_1, i_2 \in C$ and $j_1, j_2 \in \bar{C}$ satisfy 
	\[
	i_1 + j_1 = i_2 + j_2 \mod N,
	\] 
	then $(i_1-i_2, N) = (j_1-j_2,N)$, a contradiction, as $\G$ and $\G^c$ correspond to complementary sets of divisors of $N$. Thus all the sums in the sumset $C + \bar{C} =\{(i+j) \mod N \ | i \in C, j \in \bar{C}\}$ are distinct, and so $C + \bar{C}$ is of size $\mu(\G)\mu(\G^c) \leq N$. In particular, if $\mu(\G)\mu(\G^c) = N$, then $C\oplus\bar{C} = \Z_N$, and $C$ tiles $\Z_N$.
	
	Now for the given difference graph $\G$,
	\[
	\mu(\G)\mu(\G^c) = |\JJ|\mu(\G^c) = \chi(\G) \mu(\G^c) \geq N,
	\]
	and so the theorem follows.
	Note that $\mu(\G)\mu(\G^c) = N$ implies that every element of $\Z_N$ can be written uniquely as
	\[
	a = a_C + a_{\bar{C}},
	\]
where $a_C \in C$ and $a_{\bar{C}} \in \bar{C}$. We can see that the mapping $a \mapsto a_C$
is a coloring of $\G$: If $a,b$ are adjacent in $\G$ then $a_C = b_{\bar{C}}$ would imply $(a-b,N) = (a_{\bar{C}}-b_{\bar{C}},N) \in \D^c$, a contradiction. Thus we have a coloring with $|\JJ|$ colors: in other words when $\BB^\JJ$ has an orthogonal sampling set then $\mu(\G)\mu(\G^c) = N$ is equivalent to $\chi(G) = |\JJ|$.
\end{IEEEproof}

If $\G$ is perfect, for example when $N=p^M$ is a prime power, or when the hypothesis of Theorem \ref{theorem:perfect-graph} holds, then the chromatic number and clique number for $\G$ are equal, and Theorem \ref{thm:cliques -tiling} applies. Thus for prime power $N$,  if a bandlimited space has an orthogonal sampling set $\II$, then $\II$ tiles $\Z_N$. This is a well known result, see for eg \cite{coven1999tiling}, and a special case of the Fuglede's Conjecture, also known as the spectral set conjecture. This conjecture first appeared  in \cite{fuglede1974commuting} and asks, in other language,  whether the above result holds in greater generality. 

\subsection{Fuglede's conjecture} \label{section:Fuglede}

A \emph{spectral set} is a domain $\Omega \subset \mathbb{R}^N$  for which there exists a \emph{spectrum} $\{\lambda_k\}_{k \in \mathbb{Z}} \subset \mathbb{R}^N$ such that $\{e^{2\pi i \lambda_kx}\}_{k \in \mathbb{Z}} $ is an orthogonal basis for $L^2(\Omega)$. Then 

\begin{quote}
	\label{conj:fuglede}
	\emph{Conjecture} (Fuglede , \cite{fuglede1974commuting}, Spectral-Tile direction): If a domain $\Omega \subset \mathbb{R}^N$ is a spectral then it tiles $\mathbb{R}^N$.
\end{quote}

The original result of Fuglede  contained a proof of this conjecture under the assumption that $\Omega$ is a lattice subset of $\mathbb{R}^N$. Since then the conjecture has been proved to be true under more restrictive assumptions on the domain $\Omega$; for e.g. for convex planar sets \cite{iosevich2003fuglede} and union of intervals \cite{laba2001fuglede}. Further, the conjecture has been disproved in $\mathbb{R}^5$ \cite{tao2003fuglede}. For cyclic groups $\Z_N$, the conjecture is known to be true  the case when $N$ is a prime power, see for e.g. \cite{laba2002spectral}, \cite{taofugledeblog} and \cite{coven1999tiling}. The conjecture for $\Z_{p^mq}$, $p,q$ primes was proved in \cite{malikiosis2016fuglede}. Other domains where Fuglede's conjecture is known to be true includes the field of $p-$adic numbers \cite{fan2016compact, fan2015fuglede} and $\mathbb{Z}_p \times \mathbb{Z}_p$ \cite{iosevich2017fuglede}. See \cite{dutkay2014some} for the relationship between validity of conjectures in various domains. 

Thus strengthening Theorem \ref{thm:lovasz-bound} further,by proving a similar statement in terms of the chromatic number of $\G$, would prove Fuglede's conjecture (in the Spectral-Tile direction) for $\Z_N$.

\section{Conclusion and open problems}
Given a bandlimited space, we defined a difference graph such that the problem of existence and computation of orthogonal interpolating bases becomes equivalent to the problem of finding cliques in the difference graph. When $N$ is a prime power, difference graphs have nice structural properties, including perfectness, that have consequences for finding cliques and hence for orthogonal interpolation.

However, for the case when $N$ is a prime power, since orthogonal interpolation  has been otherwise well investigated, the more relevant (and interesting) situation is when $N$ is not a prime power. In this case, we provided examples of difference graphs (coming from idempotents) that are not perfect. It is not clear what perfectness implies for the corresponding idempotent, and vice versa. 

We also observed that properties of the difference  graph are closely related to tiling and spectral properties. For instance, we have not yet found a counter example to the following observation on difference graphs:

\begin{conjecture}
	\label{conj:chromatic}
	Suppose $\BB^\JJ$ has an orthogonal sampling set, and $\G$ be the associated difference graph. Then $\chi(\G) = |\JJ|$ or equivalently $\mu(\G)\mu(\G^c) = N$.
\end{conjecture}
If Conjecture \ref{conj:chromatic} is true, Fuglede's conjecture for $\Z_N$ (\cite{fuglede1974commuting}) follows in the Spectral-Tile direction via Theorem \ref{thm:cliques -tiling}.

We have also been unable to prove or provide a counter example to the following weaker version of conjecture \ref{conj:chromatic}. If $\mu(\G)\mu(\G^c) = N$ then $\mu(\G)$ divides $N$, and so we propose:

\begin{conjecture}
	\label{conj:divisibility}
	If $\G(h)$ is a difference graph constructed from an idempotent $h \in \mathbb{C}^N$, then the size of any max clique in $\G(h)$ divides $N$.
\end{conjecture} 
When $N$ is a prime power, conjecture \ref{conj:divisibility} follows trivially from Theorem \ref{thm:maximal-maximum}. For general $N$, this conjecture can be proved when $\D(h)$ is a singleton (\cite{klotz2007some}).  For arbitrary GCD-graphs, conjecture \ref{conj:divisibility} fails: consider $\G_{20}(\{2,5\})$ - this example  is from \cite{bavsic2009clique}. This GCD-graph has a maximal clique size of $6$, thus it seems to provide a counter-example to conjecture \ref{conj:divisibility}. However, an exhaustive search shows that $\G$ does not come from an idempotent, so Conjecture \ref{conj:divisibility} still remains to be resolved.


\appendices

\section{Proof of Lemma \ref{lem:h-null}}
\label{section:hnull-proof}
\begin{IEEEproof} We show that if $m \in \ZZ(h)$ and  $(r,N)=1$ then $h(mr) =0$, in other words, if $h$ vanishes at one element of an $\A_N(k) $ then it vanishes on all of $\A_N(k)$. This will prove the result.
	
	Introduce the polynomial
	\begin{equation} \label{eq:indicator-polynomial}
	p_\JJ(x) = \sum_{j \in \JJ} x^j.
	\end{equation}
	To say that $m\in \ZZ(h)$ is to say that $p_\JJ(\om_N^m)=0$. The order of $\om_N^m$ is $s=N/(m,N)$ and $\om_N^m$ is a root of the cyclotomic polynomial 
	\[
	\Phi_s(x) =\prod_{(l,s)=1}(x-\om_s^l).
	\]
	Since $\Phi_s(x)$ divides any monic polynomial that vanishes at a primitive $s$'th root of unity it divides $p_\JJ(x)$, and it follows that $p_\JJ(x)$ must vanish at all the $\omega_s^l$ with $(l,s)=1$. 
	
	Now consider evaluating $h(mr)$ for $r$ coprime to $m$. We have
	\[
	h(mr) = \frac{1}{N} \sum_{j \in \JJ} (\om_N^{mr})^j =\frac{1}{N} p_\JJ(\om_N^{mr}).
	\]
	But $\om_N^{mr}$ is also a primitive $s$'th root of unity, hence a root of $\Phi_s(x)$ and in turn a root of $p_\JJ(x)$.
\end{IEEEproof}

\bibliographystyle{IEEEtran}

\bibliography{../OIS}

\end{document}